\documentclass[reprint,twocolumn,notitlepage,prl,superscriptaddress,showpacs]{revtex4-1}
\usepackage{hyperref}
\usepackage[usenames,dvipsnames]{color}
\usepackage{amsmath}
\usepackage[english]{babel}
\usepackage{amssymb}
\usepackage{graphicx}
\usepackage{epstopdf}
\usepackage{subfig}
\usepackage{comment}
\usepackage{mathrsfs}
\usepackage{amsfonts}
\usepackage{tabularx}
\usepackage{multirow}
\usepackage[]{todonotes}
\usepackage[]{algorithm2e}

\begin{document}
%\linenumbers

\title{Optical Constants of Crystalline Bi$_2$Sr$_2$CaCu$_2$O$_{8+\delta}$ by Brillouin Light Scattering Spectroscopy}
\author{B. D. E. McNiven}
\affiliation{Department of Physics and Physical Oceanography, Memorial University of Newfoundland, St. John's, Newfoundland \& Labrador, Canada A1B 3X7} 
\author{J. P. F. LeBlanc}
\affiliation{Department of Physics and Physical Oceanography, Memorial University of Newfoundland, St. John's, Newfoundland \& Labrador, Canada A1B 3X7}
\author{G. T. Andrews}
\email{tandrews@mun.ca}
\affiliation{Department of Physics and Physical Oceanography, Memorial University of Newfoundland, St. John's, Newfoundland \& Labrador, Canada A1B 3X7} 

\date{\today}
\begin{abstract}
Room-temperature optical constants of crystalline Bi$_2$Sr$_2$CaCu$_2$O$_{8+\delta}$ were determined using data extracted from Brillouin light scattering spectra. Optical extinction coefficient-to-refractive index ratios at a wavelength of 532 nm were obtained from bulk phonon peak linewidth and frequency shift measurements and range from $0.19 \leq 2\kappa/n \leq 0.29$ for directions close to the crystallographic $c$-axis.  These ratios, and optical extinction coefficients, absorption coefficients, and imaginary parts of the dielectric function determined from these ratios and known refractive index, are in general agreement with values found in optical reflectance studies, but are 5-7 times larger than those extracted from optical interference measurements. 
\end{abstract}

\maketitle

\section{Introduction}
Bi$_2$Sr$_2$CaCu$_2$O$_{8+\delta}$ (Bi-2212) is one of three commonly studied crystalline phases of the high-$T_{c}$ superconductor Bi$_2$Sr$_2$Ca$_{n-1}$Cu$_n$O$_{2n+4+\delta}$ (BSCCO). Similar to other members of the cuprate family, Bi-2212 possesses orthorhombic symmetry and has been described as micaceous due to its layered morphology, with a cleavage plane perpendicular to the crystallographic $c$-axis \cite{Sheahen}. Bi-2212 is structurally complex, exhibiting birefringence \cite{Kobayashi} and incommensurability due to sublattice coupling along the $b$ and $c$ crystallographic axes \cite{Etrillard2001,Yurgens}.

Many properties of Bi-2212, especially the electronic and structural properties, have been studied in detail \cite{Bellini2003,Etrillard2001}, but the optical properties have received relatively little attention. Measurements of the dielectric function at visible wavelengths are scant and, with the exception of one study \cite{Bozovic}, report only the real \cite{Quijada} or imaginary \cite{Liu} part of this quantity in this region of the electromagnetic spectrum. Furthermore, there is significant variation in these values.  Two independent measurements of the complex refractive index of Bi-2212 have been reported \cite{Wang2012,Hwang} but, while the real parts of this quantity lie within 5\% of one another, the imaginary parts ({\it i.e.}, the extinction coefficients) differ by about a factor of 5.  Moreover, a third value of extinction coefficient extracted from a measurement of optical penetration depth \cite{Bozovic} is more than an order of magnitude greater than the lowest value reported in these three studies. It is therefore clear that additional studies of the optical properties of Bi-2212 are necessary to establish the true values of optical constants which appear in the literature and to broaden the knowledge base on these properties through characterization of as yet unmeasured optical constants.

In this paper, room-temperature optical extinction coefficient-to-refractive index ratios for a Bi-2212 crystal at the commonly-used wavelength of 532 nm are extracted from Brillouin spectra by measurement of bulk acoustic mode peak frequency shifts and linewidths. Knowledge of these ratios, together with published refractive index values, allows optical extinction coefficients to be determined and, consequently, other optical properties via the Kramers-Kronig transformations.

\section{Experimental Details}

\subsection{Sample}
The Bi-2212 crystal used in the present study was grown by the flux method \cite{Kanatzidis} and has a critical temperature $T_c = 78$ K. A $1\times1\times0.5$ mm$^{3}$ (001)-oriented sample for Brillouin scattering experiments was obtained from a parent crystal using mechanical exfoliation. This turned out to be somewhat challenging due to the irregular shape of the sample. 

\subsection{Brillouin Light Scattering Experiments}    
\begin{figure}[!htb]
 \centering
  \includegraphics[width=0.3\textwidth]{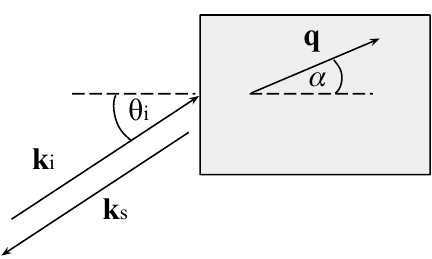}
 \caption{Backscattering geometry for Brillouin scattering experiments. {\bf{k}}$_i$ ({\bf{k}}$_s$) - incident (scattered) photon wavevector; {\bf{q}} - phonon wavevector; $\theta_i$ - angle of incidence;  $\alpha$ - direction of bulk acoustic phonon propagation as measured from the crystallographic c-axis.}
 \label{fig:Schematic}
 \end{figure}

Brillouin scattering experiments were carried out in air at room temperature using a backscattering geometry (see Fig. \ref{fig:Schematic}) with the set-up shown in ref. \cite{Andrews2007}. A single mode Nd:YVO$_4$ laser emitting at a wavelength of $\lambda_{i} = 532$ nm served as the incident light source. To minimize reflection losses, the polarization of the laser beam was rotated from vertical to horizontal by use of a half-wave plate.  It was then passed through attenuating filters to reduce the power to $\sim$10 mW and subsequently focused onto the sample using a $f=5$ cm lens with $f/\# = 2.8$. Scattered light was collected and collimated by the same lens and focused by a $f=40$ cm lens onto the entrance pinhole ($d=450$ $\mu$m) of a six-pass tandem Fabry-Perot interferometer which frequency-analyzed the scattered light.  The free spectral range of the interferometer was set to 30 GHz.  It should be noted that the use of such a low incident light power level was necessary to avoid sample damage and thermal effects caused by optical absorption; trial runs at power levels of $\geq20$ mW generated noticeable sample heating and/or damage. As a result, spectrum acquisition times of $\sim20$ hours were required and even in these circumstances only limited data could be obtained.

\section{Results and Discussion}

\subsection{Brillouin Spectra}
\begin{figure}[!htb]
 \centering

  \includegraphics[width=0.4\textwidth]{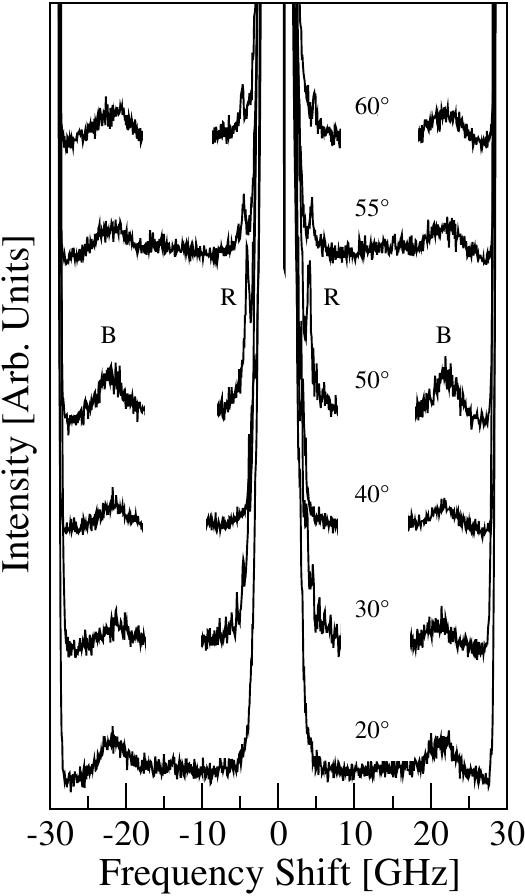}
 \caption{Room-temperature Brillouin spectra of crystalline Bi-2212 collected at incident angles ranging from $20^\circ\leq \theta_{i}\leq60^\circ$. $R$ - Rayleigh surface acoustic mode peak; $B$ - bulk acoustic mode peak.  Note: To avoid confusion, spectral regions containing a strong signal from sample mounting tape were removed.  This unwanted signal swamped features that might otherwise have been observed, including weak Brillouin peaks known to be located in this spectral region \cite{Boekholt}.}
 \label{fig:BrillouinSpectra}
 \end{figure}

Fig.\ \ref{fig:BrillouinSpectra} shows room-temperature spectra of Bi-2212 collected at incident angles $20^\circ\leq \theta_{i}\leq60^\circ$. The peak associated with the Rayleigh surface mode ($R$) was identified by the characteristic linear dependence of its frequency shift on $\sin\theta_{i}$ \cite{Sandercock1978} and its narrow linewidth relative to those of bulk mode peaks in the spectra of opaque solids. 

The peak of primary importance for this study ($B$) is due to a bulk acoustic mode.  This assignment is based upon the relatively large width of $B$ compared to that of the $R$ peak and the fact that the frequency shift is essentially independent of $\theta_{i}$ over the range probed. It is noted that while previous Brillouin studies have explicitly assigned longitudinal or transverse character to spectral peaks originating from bulk modes \cite{Boekholt}, such information is not required to extract optical constants of Bi-2212 via the method used in the present study. This is especially fortuitous given that neutron scattering studies \cite{Etrillard2001,Merritt2019} reveal the potential for multiple bulk acoustic modes (more than the usual three) in the spectrum of Bi-2212 due to its incommensurate nature, thereby complicating what is usually routine assignment of spectral peaks to either quasi-transverse or quasi-longitudinal modes. 

\begin{table*}[t]
\caption{Bulk acoustic mode frequency shifts ($f_{B}$) and associated linewidths ($\Gamma_{B}$) obtained from Brillouin spectra of crystalline Bi$_2$Sr$_2$CaCu$_2$O$_{8+\delta}$. $\theta_i$ - incident angle; $S$ ($A$) - Stokes (Anti-Stokes) scattering.  }
\begin{ruledtabular}
 \begin{tabular}{c c c c c c c }
 \multirow{1}{*}{$\theta_{i}$} & \multirow{1}{*}{$f^{S}_{B}$} & \multirow{1}{*}{$\Gamma^{S}_{B}$} & \multirow{2}{*}{$\Gamma^{S}_{B}$/$f^{S}_{B}$} & $f^{A}_{B}$ & $\Gamma^{A}_{B}$ & \multirow{2}{*}{$\Gamma^{A}_{B}$/$f^{A}_{B}$}\\
$[$deg] & [$\pm0.1$ GHz] & [GHz] & & [$\pm0.1$ GHz] & [GHz] & \\
\hline
20 & 21.6 & 4.4(2) & 0.20(1) & 21.4 & 4.6(3) & 0.21(1) \\
30 & 21.3 & 6.5(6) & 0.31(3) & 21.0 & 6.1(5) & 0.29(1) \\
40 & 21.6 & 4.2(3) & 0.19(1) & 21.6 & 4.1(2) & 0.19(1) \\
50 & 21.9 & 5.7(5) & 0.26(2) & 21.8 & 4.2(5) & 0.25(3)\\
55 & 21.6 & 5.7(3) & 0.26(1) & 21.6 & 5.7(4) & 0.26(2) \\
60 & 21.6 & 6.4(4) & 0.30(2) & 21.8 & 5.8(3) & 0.27(1)\\
\end{tabular}
\end{ruledtabular}
\label{tab:Quasi-Transverse}
\end{table*}

Table \ref{tab:Quasi-Transverse} shows the peak frequency shift ($f_{B}$) and physical linewidth ($\Gamma_{B}$) for peak $B$ at various incident angles.  Frequency shifts were obtained directly from Lorentzian fits to the raw as-collected Brillouin peaks.  The $\Gamma_{B}$ values were obtained by subtracting the instrumental width (0.28 GHz, the width of the central elastic peak) from the fitted linewidth.  This method for obtaining the physical linewidth makes use of the fact that the convolution of two Lorentzian profiles is a Lorentzian of width equal to the sum of the widths of the two constituent Lorentzians \cite{Leidecker}, and that the instrumental function and measured Brillouin peak profiles can be well-represented by Lorentzian functions \cite{Danielmeyer,Boukari}.  It also requires that the contribution of aperture broadening to the measured linewidth be negligibly small.  This was confirmed in experiments on a strongly-scattering source (acrylic) in which the difference in Brillouin linewidths obtained using collection apertures of $f/2.8$ and $f/22$ was immeasurably small. Other studies have also shown that aperture broadening is very small when a backscattering geometry is employed \cite{Gigault}. 

\subsection{Determination of Optical Constants}
The refractive index ($n$) and extinction coefficient ($\kappa$) for an opaque material are related to the bulk Brillouin peak linewidth, $\Gamma_B$, and frequency shift, $f_{B}$, through the equation
\begin{equation}
    \frac{\Gamma_B}{f_B}=\frac{2\kappa}{n},
    \label{eqn:FWHMsandercock}
\end{equation}
when $\kappa$ is large enough so that the contribution of phonon lifetime effects to the linewidth are negligible \cite{Sandercock1972_SGe}.

To obtain optical extinction coefficient-to-refractive index ratios for crystalline Bi-2212, the function 

\begin{equation}
   g(\xi) = w^{S}_{B} \left[ \frac{\Gamma^{S}_{B}}{f^{S}_{B}}-\xi \right]^2 + w^{A}_{B}\left[ \frac{\Gamma^{A}_{B}}{f^{A}_{B}}-\xi \right] ^2
   \label{eqn:minimization}
\end{equation}
was constructed, where $\xi=2\kappa/n$, and $w^{S}_{B}$ and $w^{A}_{B}$ are weighting factors equal to the reciprocals of the variances in $\Gamma^{S}_{B}/f^{S}_{B}$ and $\Gamma^{A}_{B}/f^{A}_{B}$, respectively. The value of this function is a minimum when

\begin{equation}
\xi =  \frac{w^{S}_{B} \left[ \frac{\Gamma^{S}_{B}}{f^{S}_{B}} \right] + w^{A}_{B}\left[ \frac{\Gamma^{A}_{B}}{f^{A}_{B}} \right]}{w^{S}_{B} + w^{A}_{B}}.
\label{eqn:xi}
\end{equation}
Values of $\xi = 2\kappa/n$ can then be obtained by direct substitution of pairs of $\Gamma^{i}_{B}/f^{i}_{B}$ ($i=S, A$) for a given direction from Table \ref{tab:Quasi-Transverse}  into Eq. \ref{eqn:xi}. 

Table \ref{tab:OpticalProperties} shows values of $2\kappa/n$ at 532 nm obtained using the method described above, along with derived quantities and literature values of other related optical constants.  As can be seen, the extinction coefficient-to-refractive index ratios $2\kappa/n$ obtained in the present work lie in the range $0.19 \leq 2\kappa/n \leq 0.29$ and show reasonably good agreement with those determined using $n$ and $\kappa$ values from a reflectance spectroscopy study \cite{Hwang}.  In contrast, values of these ratios extracted from the results of a combined reflectance-ellipsometry approach \cite{Bozovic} and optical interference \cite{Wang2012} are $\sim50$\% higher and 5-7 times smaller than those of the present study, respectively.

\begin{table*}[t]
\caption{Room-temperature optical properties of crystalline Bi$_2$Sr$_2$CaCu$_2$O$_{8+\delta}$ at a wavelength of 532 nm. $T_c$ - critical temperature; $n$ - refractive index; $\kappa$ - extinction coefficient; $\alpha$ - optical absorption coefficient; $d$ - optical penetration depth; $\epsilon_1$ - real part of dielectric function; $\epsilon_2$ - imaginary part of dielectric function. Entries in the ``Direction'' column specify the direction of incident light propagation inside the material, as measured from the crystallographic $c$-axis.}
\begin{ruledtabular}
 \begin{tabular}{c c c c c c c c c c}
\multirow{2}{*}{Technique} & $T_c$ & Direction & \multirow{2}{*}{$n$} &  \multirow{2}{*}{2$\kappa$/n} & \multirow{2}{*}{$\kappa$} &  $\alpha$ & $d$ & \multirow{2}{*}{$\epsilon_1$} & \multirow{2}{*}{$\epsilon_2$}\\
 & [K] & [deg] & & & &  [nm$^{-1}$] & [nm] &  \\
\hline
\multirow{6}{*}{Brillouin Spectroscopy [Present Work]} & \multirow{6}{*}{78} & 10 & & 0.21(1) & \!\!\rotatebox[origin=c]{180}{$\Rsh$} & 0.0050(2) & 202(10) & 3.96(2) & 0.84(4) \\
& & 14 & & 0.29(1) & \!\!\rotatebox[origin=c]{180}{$\Rsh$} & 0.0068(2) & 146(5) & 3.92(2) & 1.16(4) \\ 
& & 19 & & 0.19(1) & \!\!\rotatebox[origin=c]{180}{$\Rsh$} & 0.0044(2) & 229(12) & 3.96(2) & 0.76(4)\\ 
& & 23 & & 0.26(2) & \!\!\rotatebox[origin=c]{180}{$\Rsh$} & 0.0061(5) & 163(13) & 3.93(4) & 1.04(8)\\  
& & 24 & & 0.26(1) & \!\!\rotatebox[origin=c]{180}{$\Rsh$} & 0.0061(2) & 163(6) & 3.93(2) & 1.04(4)\\
& & 26 & & 0.28(1) & \!\!\rotatebox[origin=c]{180}{$\Rsh$} & 0.0066(2) & 151(5) & 3.92(2) & 1.12(4)\\
\cline{1-10}
\multirow{1}{*}{Optical Interference \cite{Wang2012}} & 85 & - & 1.9 & 0.04 & 0.04 & 0.0009\footnotemark[3] & 1060\footnotemark[1] & 3.6\footnotemark[2] & 0.15\footnotemark[2]\\
\cline{1-10}
\multirow{3}{*}{Optical Reflectance \cite{Hwang}} & 96 & \multirow{3}{*}{$\sim0$} & 1.9 & 0.24 & 0.23 & 0.005\footnotemark[3] & 184\footnotemark[1] & 3.6\footnotemark[2] & 0.87\footnotemark[2] \\ 
& 69 & & 2.0 & 0.19 & 0.19 & 0.004\footnotemark[3] & 223\footnotemark[1] & 4.0\footnotemark[2] & 0.76\footnotemark[2] \\
& 60 &  & 2.1 & 0.25 & 0.26 & 0.006\footnotemark[3] & 163\footnotemark[1] & 4.3\footnotemark[2] & 1.1\footnotemark[2]\\
\cline{1-10}
\multirow{1}{*}{Optical Reflectance-Ellipsometry \cite{Bozovic}} & - & $\sim0$ & 2.1\footnotemark[2] & 0.40 & 0.42\footnotemark[1] & 0.01\footnotemark[3] & 100 & 3.4 & 1.7\\
\end{tabular}
\end{ruledtabular}
\label{tab:OpticalProperties}
\footnotetext[1]{Estimated using $d=\lambda_{i}/4\pi\kappa$.}
\footnotetext[2]{Estimated using $\epsilon_1=n^2-\kappa^2$, $\epsilon_2=2n\kappa$.}
\footnotetext[3]{Estimated using $\alpha=1/d$.}
\end{table*}

Previous studies place the refractive index of Bi-2212 at 532 nm within 5\% of a mean value of $n=2.0$, but extinction coefficients reported in the same studies show an order of magnitude variation: $0.04 \leq \kappa \leq 0.42$ (see Table \ref{tab:OpticalProperties}) \cite{Bozovic,Wang2012,Hwang}.  Using the $2\kappa/n$ ratios of the present work and the tightly constrained refractive index of $n=2.0$ gives extinction coefficients for directions close to the $c$-axis of $0.19 \leq \kappa \leq 0.29$.  These values are in good agreement with those determined using optical reflectance methods \cite{Hwang,Bozovic}, but are $\sim7$ times larger than those obtained via optical interference \cite{Wang2012}. Similar behaviour is seen for the optical absorption coefficient ($\alpha=4\pi\kappa/\lambda_{i}$), optical penetration depth ($d=1/\alpha$), and imaginary part of the dielectric function ($\epsilon_2 = 2n\kappa$) due to the strong dependence of these quantities on $\kappa$.  On the contrary, the real part of the dielectric function $\epsilon_1 = n^2-\kappa^2$ obtained using a refractive index of $n=2.0$ and the $\kappa$ values of the present study, are similar to those of all other studies listed in Table \ref{tab:OpticalProperties} \cite{Bozovic,Hwang,Wang2012} due to the fact that $n^2 \gg \kappa^2$.

\section{Conclusion}
Brillouin light scattering spectroscopy was used to determine room temperature optical constants of crystalline Bi-2212 at a wavelength of 532 nm.  Extinction coefficient-to-refractive index ratios, and optical extinction coefficients, absorption coefficients, and imaginary parts of the dielectric function determined from these ratios and known refractive index, are consistent with values found in optical reflectance studies, but are several times larger than those obtained from optical interference measurements.  The real part of the dielectric function determined in this study is in general agreement with those of several previous studies.  

\section{Acknowledgements}
The authors would like to acknowledge Dr. J. P. Clancy at McMaster University, Canada, for supplying the samples used in this work and Dr. J. Hwang at Sungkyunkwan University, South Korea, for providing raw reflectance data for their samples.  This work was partially funded by the Natural Sciences and Engineering Council of Canada (NSERC) through Discovery Grants to Andrews (\#RGPIN-2015-04306) and LeBlanc (\#RGPIN-2017-04253).
\bibliographystyle{apsrev4-1}
\bibliography{refs.bib}

\end{document}